\documentclass[fleqn,usenatbib]{mnras}%

\usepackage[T1]{fontenc}
\usepackage{ae,aecompl}

\usepackage{graphicx}	
\usepackage{amsmath}	
\usepackage{amssymb}	
\usepackage{pdflscape}
\usepackage[dvipsnames]{xcolor}
\usepackage{longtable}

\title[The crucial role of higher-order multiplicity]{The crucial role of higher-order multiplicity in wide binary formation: A case study using the $\beta$-Pictoris moving group}

\author[P. Elliott and A. Bayo]{P. Elliott$^{1,2}$\thanks{pe210@exeter.ac.uk} 
and A. Bayo$^{3}$\\
$^1$European Southern Observatory, Alonso de Cordova 3107, Vitacura Casilla 19001, Santiago 19, Chile\\
$^2$School of Physics, University of Exeter, Stocker Road, Exeter, EX4 4QL\\
$^3$ Instituto de F\'isica y Astronom\'ia, Facultad de Ciencias, Universidad de Valpara\'iso, Av. Gran Breta\~na 1111, 5030 Casilla, Valpara\'iso, Chile}

\date{Accepted 2016 April 18. Received 2016 April 14; in original form 2016 March 15}

\pubyear{2016}

\begin{document}
\label{firstpage}
\pagerange{\pageref{firstpage}--\pageref{lastpage}}
\maketitle

\begin{abstract}
The ``in-situ'' formation of very wide binaries is hard to explain as their physical separations are beyond the typical size of a collapsing cloud core ($\approx$5000-10,000\,au). Here we investigate the formation process of such systems. We compute population statistics such as the multiplicity fraction ({\,\it MF\,}), companion-star fraction ({\,\it CSF\,}) and physical separation distribution of companions in the $\beta$-Pictoris moving group (BPMG). We compare previous multiplicity studies in younger and older regions and show that the dynamic evolution of a young population with a high degree of primordial multiplicity can lead to a processed separation distribution, similar to the field population. The evolution of outer components is attributed to the dynamical unfolding of higher-order (triple) systems; a natural consequence of which is the formation of wide binaries. We find a strong preference for wide systems to contain three or more components ($>$1000 au: 11 / 14, 10,000 au: 6 / 7). We argue that the majority of wide binaries identified in young moving groups are primordial. Under the assumption that stellar populations, within our galaxy, have statistically similar primordial multiplicity, we can infer that the paucity of wide binaries in the field is the result of dynamical evolution.

\end{abstract}




\section{Introduction}
\label{sec:intro}

It is now accepted that a significant fraction of stars are born in binary or higher-order multiple systems \citep{Duchene2013, Reipurth2014}.  However, the formation mechanism for multiple systems is still not very well understood.  The complex interplay of many physical processes leads to observed distributions of multiple systems spanning many orders of magnitude \citep[with periods, $P$, spanning the $10^{0}-10^{10}$\,day range]{Raghavan2010, Tokovinin2014b}.

In particular, the formation of very wide multiple systems ($>$10,000\,au) is hard to explain with our current theory of star formation because companions have separations beyond the limit imposed by the original hydrostatic clump size \citep[5000-10,000\,au,][]{Benson1989, Motte1998, Pineda2011}.  Therefore, in-situ formation seems extremely improbable, and either these systems are primordial but have migrated \citep{Reipurth2012}, or are non-primordial altogether (originating from different birth sites and after becoming gravitationally bound, \citealt{Kouwenhoven2010}).
To differentiate between the two scenarios we can derive statistical properties of populations and their wide binaries, i.e. \cite{Tokovinin2006, Tokovinin2015s}, and compare these with theoretical predictions of different formation mechanisms.

In this work we use the multiplicity census of the Perseus region (\citealt{Tobin2016}), the $\beta$-Pictoris moving group (hereafter, BPMG) and the field population (\citealt{Raghavan2010, Tokovinin2014a, Tokovinin2014b}) to analyse the role of higher-order multiple systems in the formation of wide binaries and population multiplicity.

Due to the long periods of wide multiple systems ($\gtrsim$30,000\,yr) detecting significant orbital motion is not possible on short ($<$100\,yr) time-scales.  {In fact we cannot directly determine whether components are gravitationally bound. }
Therefore, components are usually identified as co-moving proper motion pairs, e.g., \cite{Caballero2010}, but a high false-positive rate limits the identification to nearby sources in uncrowded fields.

In recent work \citep{Elliott2016} we identified 84 potential wide multiple systems in young nearby associations \citep{Torres2006, Torres2008, Malo2014} using 2MASS photometry, proper motions and additional spectroscopic/photometric properties. The high number of identified wide binary systems prompted the analysis of their formation presented in this paper.

\section{Sample}
\label{sec:sample}

\cite{Reipurth2012} suggested the young moving groups to be ideal populations to identify wide binaries formed by the dynamical unfolding of triple systems, thanks to their age ($\approx$10-100\,Myr) and the distance travelled from their original birth site ({ $\sim$5-10\,pc}), see Section~\ref{sec:wide_binary_justification}. From our previous work \citep{Elliott2016} we chose the BPMG in particular because of its optimal properties for wide binary analysis. The large average proper motions ($\overline{\mu_\mathrm{\alpha}}$ = 50\,mas/yr, $\overline{\mu_\mathrm{\delta}}$ = -75\,mas/yr) allowed us to derive separation limits $> $20,000\,au for a significant number (49) of systems. Due to its age \cite[21-26\,Myr, ][]{Barrado1999, Torres2006, Binks2014, Bell2015} and proximity (average distance 43\,pc) many systems (38 / 49) have been imaged previously using high-contrast techniques, looking for low-mass companions.  Additionally, almost all (42/49) systems have multi-epoch spectroscopy from works such as \cite{Malo2014} and  \cite{Elliott2014}. 

Thus, the sample studied in this paper comprises 49 single and multiple stellar systems in the BPMG.  See Table~\ref{tab:sample_targets} for full references.

\section{Companion detections and detection probabilities}
\label{sec:detec_limits}

{To study the abundance of multiple systems across a large parameter space in our sample we need both: detections of companions, and constrained parameter space for non-detections.  }
We have combined radial velocity (RV) values, high contrast imaging and direct imaging / proper motion to estimate our detection probabilities. 
Detection limits from RV measurements were derived using multi-epoch observations following \cite{Tokovinin2014a}.  For high-contrast imaging we converted the angular separation versus contrast, to physical separation versus mass-ratio, using the target's distance and the evolutionary models of \cite{Baraffe2015}. We used the detection limits described in \cite{Elliott2016}; { combining 2MASS photometry \citep{Cutri2003} and proper motions \citep[UCAC4, PPMXL, NOMAD:][]{Zacharias2012, Roeser2010, Zacharias2005}, 
for the widest parameter space ($\gtrsim$3\arcsec, $\gtrsim$130\,au).}

\begin{figure}
\begin{center}
\includegraphics[width=0.49\textwidth]{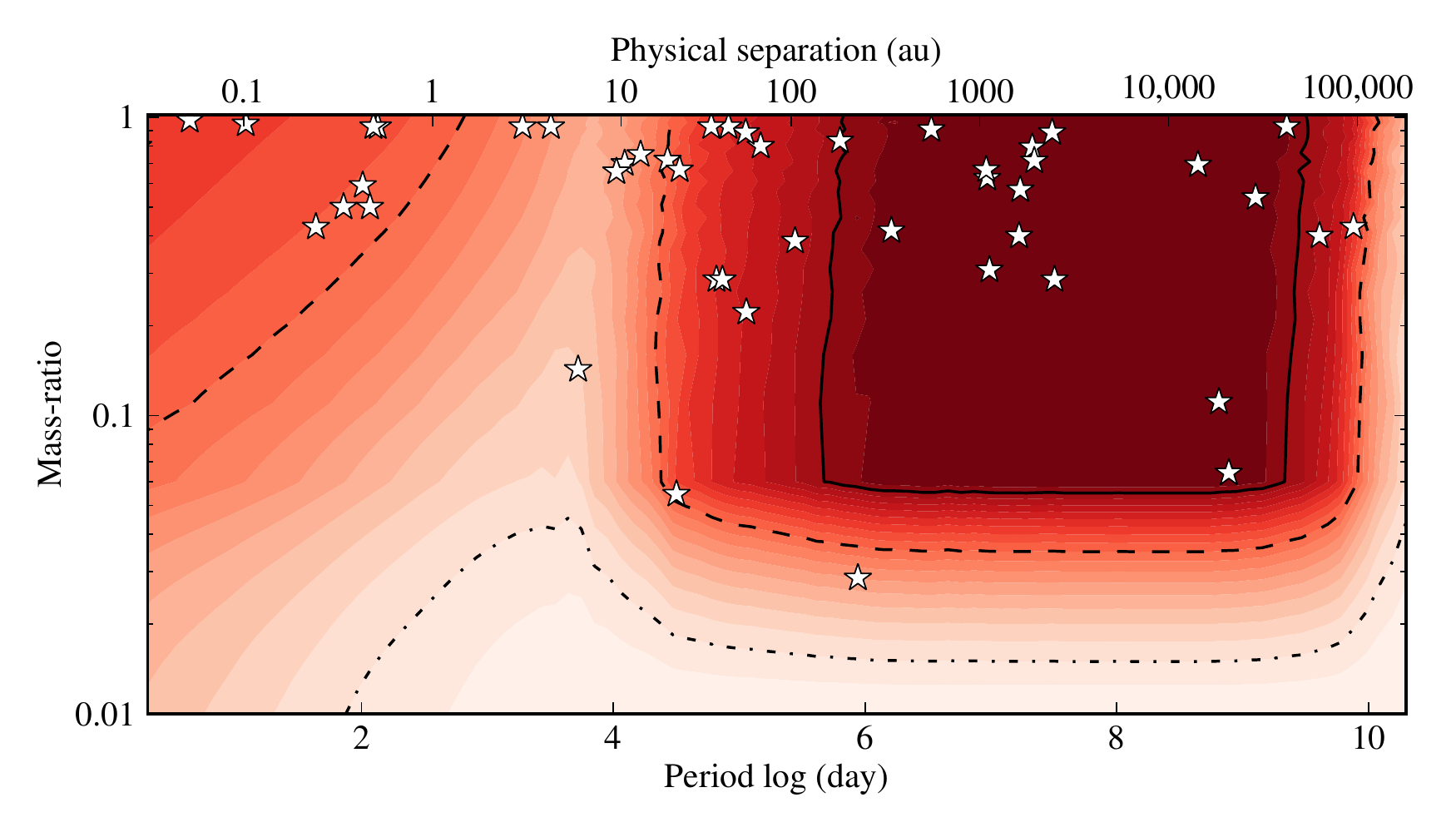}
\vspace{-0.7cm}
\caption{
Average detection probabilities (contours from red, 100\% to white, 0\%) and detected companions (white stars) in the physical separation versus mass-ratio space for the 49 systems. The solid, dashed and dash-dotted lines encompass areas with detection probabilities $\ge$  90\%, 50\% and 10\%, respectively.
}
\label{fig:all_detec_limits}
\end{center}
\end{figure}

Figure~\ref{fig:all_detec_limits} shows the average detection probabilities for our sample (contour map) in combination with the companion detections (star markers).   
Our observed multiplicity properties should not be significantly affected by our completeness, as discussed in the following section.
{ Details of the targets and the multiple systems studied in this work can be found in Tables~\ref{tab:sample_targets} and \ref{tab:mms_sample}, respectively.}

\section{Detection completeness}
\label{sec:completeness}

We did not attempt to {\it correct} our statistics with our detection probabilities, { however,} below we justify why this should not significantly affect our results.  

In this analysis before averaging the detection probabilities over the sample, as in Figure~\ref{fig:all_detec_limits}, we first consider whether a companion is likely to inhabit the parameter space based on stability arguments. \cite{Tokovinin2014b} showed that the region of instability (for triple and higher-order systems), based on the ratio of periods in the system is between certain values ($4.7 < P_L / P_S < 47$).  We use the mid-value (26) of this criterion to mask regions of detection probability space using previously identified companions in the system.  For example, if a system has a companion with a period of 100\,day then we mask the region 4--2600\,day.  Masking in this case means setting a detection probability of 1.  This is because we use these probabilities to predict where the {\it missing} companions would be and we have indirectly probed this space with our stability arguments.  We repeated this for all objects to create a new set of detection probability arrays.  We then averaged these detection probabilities over the mass-ratio in each period bin because there is no significant evidence that the mass-ratio is a function of period \citep{Reggiani2013, Tokovinin2014b}.

\begin{figure}
\begin{center}
\includegraphics[width=0.49\textwidth]{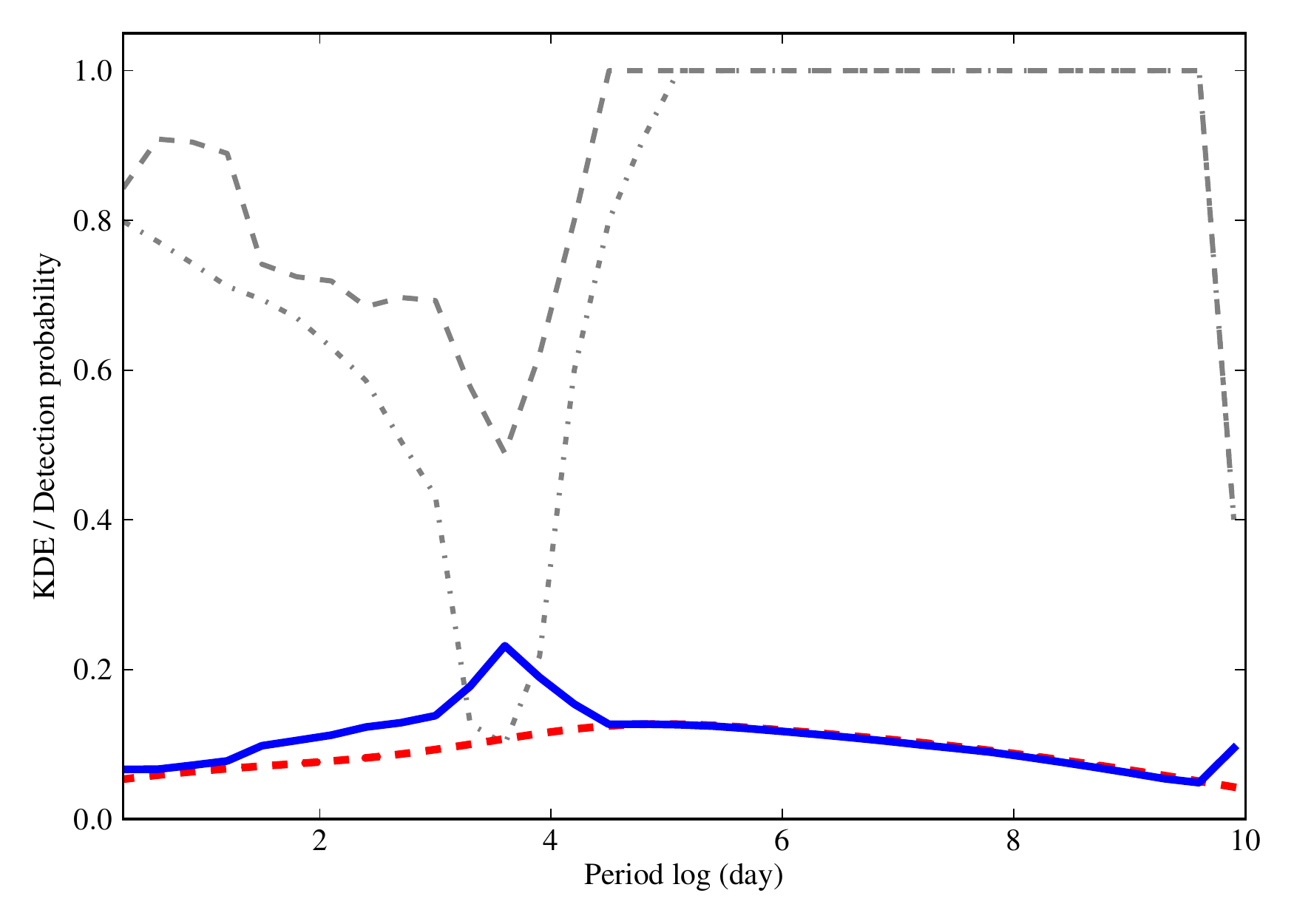}
\vspace{-0.2cm}
\caption{Kernel density estimations for the period versus companion-star fraction.  The dashed and dash-dotted grey lines represent median detection probabilities (averaged over mass-ratio) with and without dynamical stability considerations, respectively.  The red dashed line is the raw sample analysed in the body of this work and the blue line is the detection-probability {\it corrected} distribution.}
\label{fig:detection_correc}
\end{center}
\end{figure}

Figure~\ref{fig:detection_correc} shows our results based on the period distribution of systems in our sample.  We used kernel density estimations (KDEs) to perform this analysis which reduces the phase effect when binning data, see \cite{scott2009multivariate} for details.  

The grey dotted and dash-dotted lines are the median detection probabilities with and without dynamical constraints, respectively.  This Figure shows that although we are very insensitive in regions $\approx$4\,day, due to identified companions in the systems, the likely number of missing companions is not very significant. The red dashed line and blue line are the KDEs of the raw sample and the {\it corrected} sample (KDE / detection probability), respectively. We then integrated these curves to calculate the quantitative difference in {\it CSF} values.  The difference was $\approx$15\% i.e. the derived {\it CSF} value derived in this work is most likely $\approx$1.17 as opposed to 1.02. This increase would not significantly affect any conclusions presented in this work, in fact this value would put our results in even better agreement with the Class 0 {\it CSF} value from \cite{Tobin2016}, { see Section~\ref{sec:mf_comparison} for details.}

{
\onecolumn
\begin{landscape}
\tiny
\LTcapwidth=\textwidth
\begin{longtable}{p{0.9cm} p{0.5cm} p{4cm} p{1.2cm} p{1.5cm} p{0.8cm} p{0.8cm} p{0.8cm} p{1.cm} p{1.3cm} p{3.0cm}  p{2.5cm}}
\caption{Observtional properties of $\beta$-Pictoris targets studied in this work.} \label{tab:sample_targets}\\
\hline\hline\\
ID &  Comp.$^a$ & Common ID & R.A. & DEC. & Dist. & V & K & Mem. ref. & 2MASS lim. & AO refs. & RV refs.\\
 &   &   &  hh:mm:ss.s & dd:mm:ss & (pc) & (mag.) & (mag.) & & (\arcsec) &  & \\
\hline\\
\endfirsthead
\hline\hline\\
ID &  Comp. & Common ID & R.A. & DEC. & Dist. & V & K & Mem. ref. & 2MASS lim. & AO refs. & RV refs.\\
 &   &   &  hh:mm:ss.s & dd:mm:ss & (pc) & (mag.) & (mag.) & & (\arcsec) &  & \\
\hline\\
\endhead
BPC 10   &   A  &   V1005 Ori   &   04:59:34.8   &   +01:47:01.0   &   			25.9   &   10.11   &   6.26   &   28   &   4000   &   3, 2, 8, 12, 17, 19, 24   &   9, 1, 12 \\
BPC 10   &   B  &   2MASS J05015665+0108429   &   05:01:56.7   &   +01:08:43.0   &   			25.9   &   12.87   &   7.68   &   11   &   4000   &      &    \\
BPC 12   &   A  &   HIP 23418aa   &   05:01:58.8   &   +09:58:59.0   &   			33.2   &   11.67   &   6.37   &   20, 28   &   1000   &   10   &    \\
BPC 14   &   A  &   BD-21 1074b   &   05:06:49.5   &   -21:35:04.0   &   			20.2   &   11.30   &   6.11   &   20, 28   &   6000   &   8   &   12, 9, 12 \\
BPC 16   &   B  &   2MASS J05200029+0613036   &   05:20:00.3   &   +06:13:04.0   &   			67.8   &   11.42   &   8.57   &   28   &   600   &   10   &   9 \\
BPC 16   &   A  &   RX J0520.5+0616   &   05:20:31.8   &   +06:16:11.0   &   			71.0   &   11.64   &   8.57   &   28   &   600   &   10   &   9 \\
BPC 16   &   C  &   2MASS J05195327+0617258   &   05:19:53.3   &   +06:17:26.0   &   			71.0   &   18.31   &   12.41   &   11   &   600   &      &    \\
BPC 19   &   A  &   Beta Pic   &   05:47:17.1   &   -51:03:59.0   &   			19.4   &   3.85   &   3.53   &   28   &   4500   &   7, 26, 3, 29   &    \\
BPC 2   &   A  &   GJ 3076   &   01:11:25.4   &   +15:26:21.0   &   			19.3   &   13.96   &   8.21   &   20   &   5000   &      &   20 \\
BPC 22   &   A  &   TWA22   &   10:17:26.9   &   -53:54:27.0   &   			17.6   &   13.89   &   7.69   &   28   &   5500   &   12, 7, 2   &   9, 12 \\
BPC 26   &   A  &   V824 Ara   &   17:17:25.5   &   -66:57:04.0   &   			31.4   &   6.86   &   4.70   &   28   &   3100   &   23, 7, 2, 19, 24   &   25 \\
BPC 26   &   B  &   HD 155555 c   &   17:17:31.3   &   -66:57:06.0   &   			31.4   &   12.89   &   7.63   &   28   &   3100   &   3   &    \\
BPC 28   &   A  &   GSC 08350-01924   &   17:29:20.7   &   -50:14:53.0   &   			66.3   &   12.96   &   7.99   &   20, 28   &   350   &   10   &    \\
BPC 3   &   A  &   Barta 161 12   &   01:35:13.9   &   -07:12:52.0   &   			37.9   &   13.43   &   8.08   &   20, 16   &   2500   &      &   20 \\
BPC 3   &   B  &   2MASS J01354915-0753470   &   01:35:49.2   &   -07:53:47.0   &   			37.9   &   13.79   &   9.81   &   11   &   2500   &      &    \\
BPC 30   &   A  &   HD 161460   &   17:48:33.7   &   -53:06:43.0   &   			71.0   &   8.99   &   6.78   &   28   &   600   &   10, 7   &   22 \\
BPC 30   &   B  &   2MASS J17483374-5306118   &   17:48:33.8   &   -53:06:12.0   &   			71.0   &   13.65   &   9.27   &   11   &   600   &      &    \\
BPC 31   &   A  &   HD 164249   &   18:03:03.4   &   -51:38:56.0   &   			48.1   &   7.01   &   5.91   &   28   &   2000   &   7, 26   &   12, 9, 11, 12 \\
BPC 31   &   C  &   2MASS J18011138-5125594   &   18:01:11.3   &   -51:26:00.0   &   			48.1   &   14.79   &   11.27   &   11   &   2000   &      &    \\
BPC 36   &   A  &   2MASS J18420694-5554254    &   18:42:06.9   &   -55:54:25.0   &   			51.9   &   13.54   &   8.58   &   20   &   500   &      &   20 \\
BPC 36   &   B  &   2MASS J18420483-5554126   &   18:42:04.8   &   -55:54:13.0   &   			51.9   &   15.11   &   9.85   &   11   &   500   &      &    \\
BPC 37   &   A  &   HD 172555   &   18:45:26.9   &   -64:52:17.0   &   			28.5   &   4.80   &   4.30   &   28   &   3500   &   7, 3, 2, 29, 26, 24   &    \\
BPC 37   &   B  &   CD-64 1208   &   18:45:37.0   &   -64:51:46.0   &   			28.5   &   9.97   &   6.10   &   28   &   3500   &   24, 7, 2, 19   &   22 \\
BPC 39   &   B  &   Smethells 20   &   18:46:52.6   &   -62:10:36.0   &   			52.4   &   12.20   &   7.85   &   20, 28   &   2000   &   10, 3   &   27, 9 \\
BPC 39   &   A  &   2MASS J18480637-6213470   &   18:48:06.5   &   -62:13:47.0   &   			52.4   &   7.29   &   6.14   &   11   &   2000   &      &    \\
BPC 4   &   B  &   HIP 10679   &   02:17:24.7   &   +28:44:30.0   &   			27.3   &   7.78   &   6.26   &   28   &   2500   &   3, 5   &   25 \\
BPC 4   &   A  &   HD 14082   &   02:17:25.3   &   +28:44:42.0   &   			34.5   &   7.04   &   5.79   &   28   &   2500   &      &   25 \\
BPC 4   &   C  &   2MASS J02160734+2856530   &   02:16:07.3   &   +28:56:53.0   &   			34.5   &   15.74   &   11.61   &   11   &   2500   &      &    \\
BPC 41   &   A  &   PZ Tel   &   18:53:05.9   &   -50:10:50.0   &   			51.5   &   8.42   &   6.37   &   28   &   1900   &   3, 24, 19   &   12, 12 \\
BPC 42   &   A  &   TYC 6872-1011-1   &   18:58:04.2   &   -29:53:05.0   &   			82.6   &   11.73   &   8.02   &   20, 28   &   300   &   10   &   11 \\
BPC 42   &   B  &   2MASS J18580464-2953320   &   18:58:04.6   &   -29:53:32.0   &   			82.6   &   12.85   &   8.76   &   11   &   300   &      &    \\
BPC 43   &   A  &   CD-26 13904   &   19:11:44.7   &   -26:04:09.0   &   			78.9   &   10.24   &   7.37   &   28   &   600   &   10   &   9, 11 \\
BPC 44   &   A  &   Eta Tel   &   19:22:51.2   &   -54:25:26.0   &   			48.2   &   5.17   &   5.01   &   28   &   2000   &   23, 18, 3, 6, 29   &    \\
BPC 44   &   B  &   HD 181327   &   19:22:58.9   &   -54:32:17.0   &   			51.8   &   7.05   &   5.91   &   28   &   2000   &   7, 3   &   9, 12 \\
BPC 47   &   B  &   UCAC3 116-474938   &   19:56:02.9   &   -32:07:19.0   &   			58.4   &   13.27   &   8.11   &   20, 28   &   1800   &   3, 4   &   27, 20, 11 \\
BPC 47   &   A  &   TYC 7443-1102-1   &   19:56:04.4   &   -32:07:38.0   &   			56.3   &   11.78   &   7.85   &   20, 28   &   1800   &   8, 3, 4   &   27, 9, 11 \\
BPC 52   &   C  &   AU Mic   &   20:45:09.5   &   -31:20:27.0   &   			9.8   &   8.80   &   4.53   &   28   &   10000   &   23, 15, 2, 8, 17, 14, 3, 5, 24   &   12, 9, 1, 12 \\
BPC 52   &   B  &   AT Mic S   &   20:41:51.1   &   -32:26:10.0   &   			9.5   &   10.89   &   4.94   &   28   &   10500   &   24, 2, 19   &   1, 12 \\
BPC 52   &   A  &   AT Mic N   &   20:41:51.2   &   -32:26:07.0   &   			10.2   &   10.89   &   4.94   &   28   &   10500   &   23, 15, 2, 3, 19, 24   &   1, 12 \\
BPC 55   &   A  &   2MASS J20434114-2433534 A   &   20:43:41.1   &   -24:33:53.0   &   			45.1   &   14.45   &   7.76   &   20   &   2200   &      &   20 \\
BPC 55   &   B  &   2MASS J20434114-2433534 B   &   20:43:41.1   &   -24:33:53.0   &   			45.1   &   14.45   &   7.76   &   28   &   2200   &      &   20 \\
BPC 57   &   A  &   HD 199143   &   20:55:47.7   &   -17:06:51.0   &   			45.7   &   7.33   &   5.81   &   28   &   1200   &   23, 21   &    \\
BPC 57   &   B  &   AZ Cap   &   20:56:02.7   &   -17:10:54.0   &   			45.7   &   10.62   &   7.04   &   20, 28   &   1500   &   23, 24, 19   &   9 \\
BPC 6   &   B  &   BD+30 397 b   &   02:27:28.1   &   +30:58:41.0   &   			40.0   &   12.39   &   7.92   &   28   &   2400   &   5   &   11 \\
BPC 6   &   A  &   AG Tri   &   02:27:29.3   &   +30:58:25.0   &   			40.0   &   10.15   &   7.08   &   28   &   2400   &   5   &   11, 1 \\
BPC 63   &   A  &   WW PsA   &   22:44:58.0   &   -33:15:02.0   &   			23.6   &   12.10   &   6.93   &   28   &   4800   &   8   &   9, 20, 1 \\
BPC 63   &   B  &   TX PsA   &   22:45:00.0   &   -33:15:26.0   &   			20.1   &   13.35   &   7.79   &   28   &   4800   &   8, 10, 3   &   9, 1 \\
BPC 65   &   C  &   G 271-110   &   01:36:55.1   &   -06:47:38.0   &   			29.5   &   \ldots   &   8.86   &   20   &   4100   &   4   &    \\
BPC 65   &   B  &   2MASS J01373545-0645375   &   01:37:35.4   &   -06:45:38.0   &   			24.0   &   7.68   &   5.75   &   11   &   4100   &   14, 24, 17   &   12 \\
BPC 65   &   A  &   2MASS J01334282-0701311   &   01:33:42.8   &   -07:01:31.0   &   			29.5   &   5.77   &   4.26   &   11   &   4100   &      &   27 \\
BPC 68   &   A  &   2MASS J14142141-1521215   &   14:14:21.4   &   -15:21:21.5   &   			30.2   &   10.39   &   6.60   &   20   &   3200   &   7   &   27, 20 \\
BPC 68   &   B  &   2MASS J14141700-1521125   &   14:14:17.0   &   -15:21:13.0   &   			30.2   &   15.60   &   8.82   &   11   &   3200   &      &    \\
BPC 69   &   B  &   2MASS J21212873-6655063   &   21:21:28.7   &   -66:55:06.3   &   			30.2   &   10.66   &   7.01   &   20   &   3200   &      &    \\
BPC 69   &   A  &   2MASS J21212446-6654573   &   21:21:24.5   &   -66:54:57.0   &   			30.2   &   9.00   &   6.40   &   11   &   3200   &   15   &    \\
BPC 70   &   A  &   2MASS J20100002-2801410   &   20:10:00.0   &   -28:01:41.0   &   			48.0   &   15.13   &   7.73   &   20   &   2700   &   4   &   20 \\
BPC 70   &   B  &   2MASS J20085122-2740536   &   20:08:51.1   &   -27:40:54.0   &   			48.0   &   16.27   &   11.78   &   11   &   2700   &      &    \\
BPC 8   &   A  &   EXO 0235.2-5216   &   02:36:51.7   &   -52:03:04.0   &   			28.6   &   12.07   &   7.50   &   20, 28   &   3500   &   23, 6   &   27, 20 \\
BPC 9   &   A  &   HD 29391   &   04:37:36.1   &   -02:28:25.0   &   			29.4   &   5.22   &   4.54   &   28   &   3400   &   26, 3, 13   &    \\
BPC 9   &   B  &   2MASS J04373746-0229282   &   04:37:37.5   &   -02:29:28.0   &   			29.4   &   10.64   &   6.41   &   11   &   3400   &   8, 15   &   27, 9, 1, 12 \\
\hline\\[1ex]
\end{longtable}
{\cite{Bailey2012}: 1, \cite{Biller2007}: 2, \cite{Biller2013}: 3, \cite{Bowler2015}: 4, \cite{Brandt2014}: 5, \cite{Chauvin2003}: 6,
\cite{Chauvin2010}: 7, \cite{Delorme2012}: 8, 
\cite{Elliott2014}: 9, \cite{Elliott2015}: 10, 
\cite{Elliott2016}: 11, Elliott et al. in prep.: 12, 
\cite{Heinze2010}: 13, \cite{Janson2013}: 14,\cite{Kasper2007}: 15, \cite{Kraus2014}: 16,
\cite{Lafreniere2007}: 17, \cite{Lowrance2005}: 18,
\cite{Masciadri2005}: 19, \cite{Malo2014}: 20,
\cite{Metchev2009}: 21, \cite{Messina2010}: 22, 
\cite{Neuhauser2003}: 23,\cite{Nielsen2010}: 24,
\cite{Nordstrom2004}: 25, \cite{Rameau2013}: 26,
\cite{Kordopatis2013}: 27,\cite{Torres2006}: 28,
\cite{Vigan2012}: 29}
\noindent{$^a${The "ID" column + "Comp." column is a unique identifier for resolved targets.  This unique identifier links this Table to Table~\ref{tab:mms_sample}, this avoids any confusion for targets with multiple unresolved components.}}
\end{landscape}}

\twocolumn

\section{Defining wide multiple systems in the young moving groups}
\label{sec:wide_binary_justification}

Wide components are usually identified from the galactic motion (the quantity used, at least in initial identification, is proper motion) they share with their associated primary.  With the proper motion, radial velocity and either a photometric or kinematic distance, the 3D kinematics can be derived.  However, young moving group members are also classified on this basis, i.e. looking for collections of objects sharing the same galactic motion. Therefore we need to clarify what is a component in a multiple system and what is {\it another} moving group member.

To investigate this we looked at the different scales of physical clustering within the BPMG.  We cannot use projected separations from spatial positions for this analysis as the proximity of the stars means the group has significant depth (in galactic co-ordinates: X$\approx$160, Y$\approx$100, Z$\approx$45\,pc).  We can, however, use the derived photometric distances to known members and perform queries in a 3D spatial volume. This is similar to the analysis presented in \cite{Larson1995} where the authors identified a break in the density of objects as a function of angular separation in Taurus (58\arcsec, $\approx$8000\,au). However, as the depth of Taurus is significantly lower ($\approx$20\,pc) an angular query was sufficient, assuming a fixed distance to all sources.

\begin{figure}
\begin{center}
\includegraphics[width=0.49\textwidth]{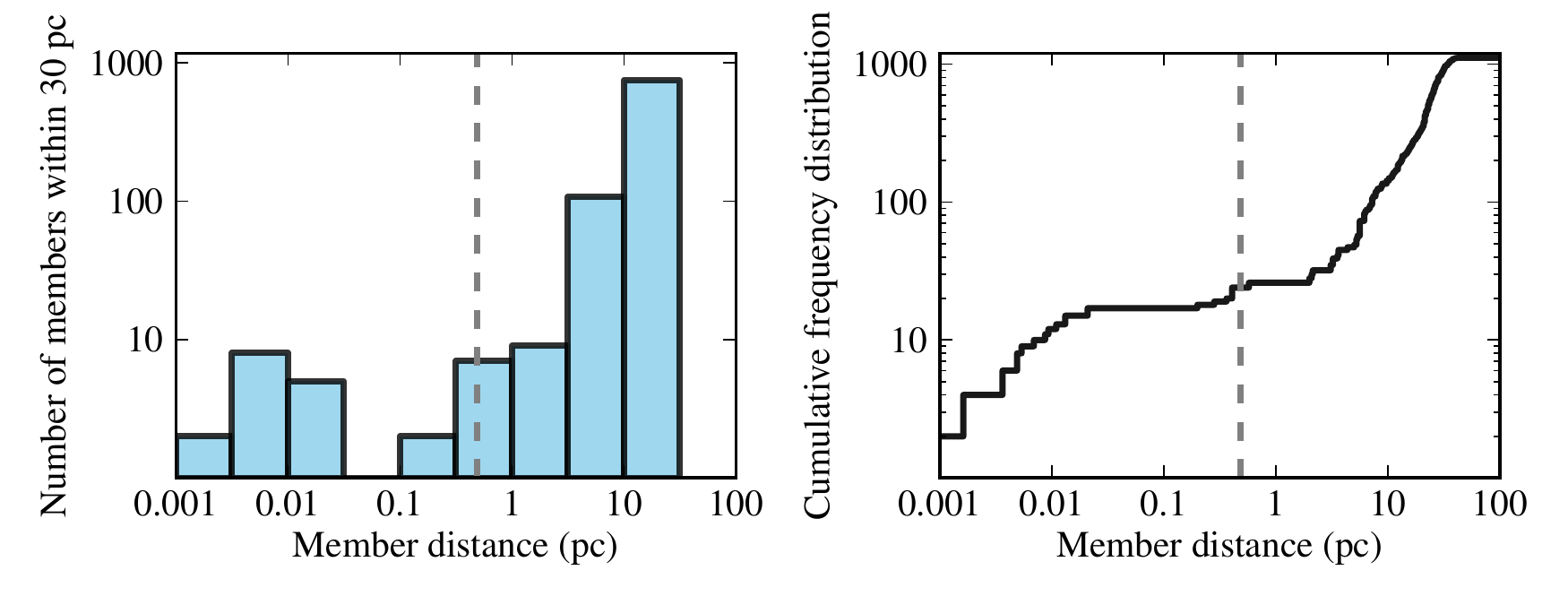}
\vspace{-0.2cm}
\caption{{\it Left panel}: The number of members within a 30 x 30 x 30\,pc volume of space for each member of the BPMG, summed over the population as a function of the physical separation. {\it Right panel}:}
\label{fig:nearest_neighbour}
\end{center}
\end{figure}

For our analysis we queried a 30\,pc volume around each bona-fide member of the BPMG using the derived $X$, $Y$ and $Z$ galactic positions. We calculated the magnitude of the distance difference $\sqrt{(X_1-X_2)^2 + (Y_1-Y_2)^2 + (Z_1-Z_2)^2}$ between the considered bona-fide member and any other member in this volume.  For wide components that share statistically similar radial velocities and proper motions and without $X$, $Y$ and $Z$ values, the distance magnitude is computed from the angular separation of the components and the distance to the bona-fide member.  We collected all of these distance magnitudes together and analysed the resultant distribution, shown in Figure~\ref{fig:nearest_neighbour}. Note that the sum of separation bins ($\approx$900) is much larger than the input sample (70) because there is a lot of overlapping parameter space from query to query (for example if all 70 objects had 10 other members within the 30\,pc volume this would result in a sum of 700 objects in the distribution).

In Figure~\ref{fig:nearest_neighbour} we see that there are two components in the distribution.  We also show the 100,000\,au limit ($\approx$0.5\,pc), the widest companions we are sensitive to from the limits derived in \cite{Elliott2016}.  We do not attempt to fit any power-laws to the components observed in the right panel of Figure~\ref{fig:nearest_neighbour} as in \cite{Larson1995}.  The BPMG sample is far from complete and there are likely still many lower-mass members to be identified (similar to the Tuc-Hor moving group, see \citealt{Kraus2014}).  Additionally our detection limit of 100,000\,au means at this point we are no longer sensitive to identifying wide companions and therefore any analysis of the resultant distribution is problematic.

We use Figure~\ref{fig:nearest_neighbour} to demonstrate that there are at least two different spatial scales (breaking at $\approx$1-5\,pc), demonstrated clearly in the right panel.  This break in the typical spatial density of members in BPMG is evidence that the population of identified wide companions (100,000\,au and smaller) is not just a continuation of clustering down to smaller scales.  These wide companions form a distinct population, their separation distribution is a result of a different physical process. 

\section{Results}
\label{sec:result}

In the following section we present analysis using the multiplicity fraction ({\it MF}) defined by
{
\begin{equation}
\label{eq:mf}
MF=\frac{B+T+Qu+Qi+Se...}{S+B+T+Qu+Qi+Se...}
\end{equation}}

and the companion-star fraction ({\it CSF}), defined by
{
\begin{equation}
CSF=\frac{B+2T+3Qu+4Qi+5Se...}{S+B+T+Qu+Qi+Se...}
\end{equation}}
where S, B, T, Qu, Qi, Se represent the number of single, binary, triple, quadruple, quintuple and sextuple systems, respectively. 

Additionally we use the physical separation (a proxy for the semi-major axis) versus {\it CSF} which allows us to investigate the dynamical evolution of multiple systems.
{
\begin{table*}
\caption{Multiple systems used in this work.}
\begin{tabular}{p{1.5cm} p{1cm} p{2cm} p{1cm} p{1.2cm} p{1.2cm} p{1cm} p{1cm} p{1.0cm}}
\hline\hline\\
ID &  Comp. & Struc. & Hier. & M1 & M2 & Sep. & Type$^a$ & Ref. \\
&  & &  & (M$_\odot$) & (M$_\odot$) & (au)  &  \\
\hline\\
BPC 10  &    B   &   A, B, *   &   L1   &   0.70   &   0.30   &   81044   &   Cpm   &   EL16 \\
BPC 10  &  A	&   Aa, Ab, A   &   L11   &   0.70   &   0.30   &   0.26   &   s   &   EL14 \\
BPC 12  &    A   &   A, B, *   &   L1   &   0.50   &   0.40   &   46   &   v   &   HIP \\
BPC 12  &  A	&   Aa, Ab, A   &   L11   &   1.00   &   0.95   &   0.10   &   S   &   DE99 \\
BPC 14  &    B   &   A, B, *   &   L1   &   0.60   &   0.50   &   122   &   v   &   WDS \\
BPC 14  &    A   &   Ba, Bb, B   &   L12   &   0.30   &   0.20   &   13   &   v   &   WDS \\
BPC 16  &    C   &   A, B, *   &   L1   &   0.70   &   0.70   &   39269   &   Cpm   &   EL16 \\
BPC 16  &    A   &   Aa, Ab, A   &   L11   &   0.70   &   0.20   &   28   &   v   &   EL15 \\
BPC 19  &  A	&   A, B, *   &   L1   &   1.40   &   0.01   &   7.78   &   v   &   WDS \\
BPC 2  &  A	&   A, B, *   &   L1   &   0.10   &   0.07   &   5.79   &   v   &   WDS \\
BPC 22  &  A	&   A, B, *   &   L1   &   0.10   &   0.10   &   1.76   &   v   &   WDS \\
BPC 26  &    B   &   A, B, *   &   L1   &   1.30   &   0.40   &   1065   &   v   &   AF15 \\
BPC 26  &  A	&   Aa, Ab, A   &   L11   &   1.30   &   1.04   &   0.94   &   s   &   WDS \\
BPC 28  &    A   &   A, B, *   &   L1   &   0.63   &   0.56   &   48   &   v   &   EL15 \\
BPC 3  &  A	&   Aa, Ab, A   &   L1   &   0.20   &   0.20   &   0.38   &   s   &   MA14 \\
BPC 30  &  A	&   Aa, Ab, A   &   L1   &   1.30   &   1.27   &   0.71   &   s   &   ME10 \\
BPC 31  &    B   &   Aa, Ab, A   &   L1   &   1.20   &   0.50   &   323   &   v   &   WDS \\
BPC 36  &    B   &   A, B, *   &   L1   &   0.50   &   0.20   &   1138   &   Cpm   &   WDS \\
BPC 37  &    B   &   A, B, *   &   L1   &   0.50   &   0.70   &   2035   &   v   &   WDS \\
BPC 37  &  B	&   Ba, Bb, B   &   L12   &   0.70   &   0.10   &   5.71   &   v   &   WDS \\
BPC 37  &  B	&   Ba1, Ba2, Ba   &   L121   &   0.70   &   0.70   &   0.86   &   s   &   ME10 \\
BPC 39  &    A   &   A, B, *   &   L1   &   1.30   &   0.70   &   28894   &   v   &   AF15 \\
BPC 39  &  A	&   Aa, Ab, A   &   L11   &   1.30   &   0.65   &   1.57   &   s   &   MO13 \\
BPC 4  &    B   &   Aa, Ab, A   &   L1   &   1.10   &   1.00   &   390   &   v   &   WDS \\
BPC 41  &    A   &   A, B, *   &   L1   &   1.10   &   0.06   &   21   &   v   &   WDS \\
BPC 41  &  A	&   Aa, Ab, A   &   L11   &   1.10   &   0.65   &   1.54   &   s   &   NO04 \\
BPC 42  &    B   &   A, B, *   &   L1   &   0.90   &   0.80   &   2290   &   Cpm   &   AF15 \\
BPC 43  &    A   &   A, B, *   &   L1   &   1.03   &   0.74   &   21   &   v   &   EL15 \\
BPC 44  &    B   &   A, B, *   &   L1   &   1.40   &   0.09   &   20072   &   Cpm   &   WDS \\
BPC 44  &    B   &   Aa, Ab, A   &   L11   &   1.40   &   0.04   &   203   &   v   &   WDS \\
BPC 47  &    B   &   A, B, *   &   L1   &   0.70   &   0.40   &   1488   &   Cpm   &   WDS \\
BPC 47  &    B   &   Ba, Bb, B   &   L12   &   0.40   &   0.30   &   11   &   v   &   WDS \\
BPC 47  &  B	&   Ba1, Ba2, Ba   &   L121   &   0.40   &   0.40   &   0.56   &   s   &   MA14 \\
BPC 52  &    C   &   A, B, *   &   L1   &   0.50   &   0.20   &   44377   &   Cpm   &   EL16 \\
BPC 52  &    C   &   Ba, Bb, B   &   L12   &   0.20   &   0.20   &   27   &   Cpm   &   WDS \\
BPC 55  &  B	&   A, B, *   &   L1   &   0.60   &   0.60   &   4.51   &   v   &   MA13 \\
BPC 55  &  A	&   Aa, Ab, A   &   L11   &   0.60   &   0.30   &   1.35   &   s   &   ELPP \\
BPC 57  &    B   &   A, B, *   &   L1   &   1.30   &   0.90   &   14764   &   Cpm   &   WDS \\
BPC 57  &    A   &   Aa, Ab, A   &   L11   &   0.90   &   0.20   &   50   &   v   &   WDS \\
BPC 57  &    B   &   Ba, Bb, B   &   L12   &   1.30   &   0.50   &   100   &   v   &   WDS \\
BPC 6  &    B   &   A, B, *   &   L1   &   0.80   &   0.50   &   894   &   Cpm   &   AF15 \\
BPC 63  &    B   &   A, B, *   &   L1   &   0.30   &   0.20   &   681   &   Cpm   &   SO02 \\
BPC 65  &    C   &   Ba, Bb, B   &   L12   &   0.90   &   0.10   &   14635   &   Cpm   &   WDS \\
BPC 68  &    B   &   A, B, *   &   L1   &   0.70   &   0.20   &   1905   &   Cpm   &   WDS \\
BPC 68  &    A   &   Aa, Ab, A   &   L11   &   0.70   &   0.20   &   33   &   v   &   WDS \\
BPC 69  &  A	&   Aa, Ab, A   &   L1   &   0.80   &   0.80   &   0.30   &   s   &   TO06 \\
BPC 70  &    A   &   Aa, Ab, A   &   L1   &   0.70   &   0.70   &   34   &   v   &   WDS \\
BPC 8  &  A	&   Aa, Ab, A   &   L1   &   0.40   &   0.20   &   0.29   &   s   &   ELPP \\
BPC 9  &    B   &   A, B, *   &   L1   &   1.40   &   1.11   &   1957   &   Cpm   &   FE06 \\
BPC 9  &  B	&   Ba, Bb, B   &   L12   &   0.67   &   0.44   &   8.83   &   V   &   MO15 \\
\hline\\[-0.1ex]
\end{tabular}\\
{$^a${Cpm: Common proper motion companion, s: spectroscopic binary without orbital solution, S: spectroscopic binary with orbital solution, v: visual binary without orbital solution, V: visual binary with orbital solution.}}
{HIP: \cite{Perryman1997}, DE99: \cite{Delfosse1999}, WDS: \cite{Mason2001}, SO02: \cite{Song2002}, NO04: \cite{Nordstrom2004}, \cite{Torres2006}, ME10: \cite{Messina2010}, BA12: \cite{Bailey2012}, MA13: \cite{Malo2013}, MA14: \cite{Malo2014}, EL14: \cite{Elliott2014}, EL15: \cite{Elliott2015}, AF15: \cite{Alonso-Floriano2015}, MO15: \cite{Montet2015}, EL16: \cite{Elliott2016}, ELPP: Elliott et al. in prep.}
\label{tab:mms_sample}
\end{table*}
}

\subsection{Linking multiplicity: young - old populations}
\label{sec:mf_comparison}

The {\it MF} and {\it CSF} of the BPMG, shown in row four of Table~\ref{tab:mf_summary}, are incompatible with that of the older field population, at $>5 \sigma$ level.
In contrast, our derived quantities are compatible with those of extremely young Class 0 sources from the work of \cite{Chen2013} ($a\approx$50-5000\,au) and \cite{Tobin2016} ($a\approx$15-10,000\,au).  However, we are sensitive to a much wider range of physical separations.  Therefore, the apparent conservation of the {\it MF} and {\it CSF} values between these regions can be interpreted as an evolution of the separation distribution.  Assuming all considered regions have statistically similar primordial multiple system populations \citep{Kroupa2011}, we investigate if significant dynamical evolution from Class 0 (Perseus), to Class I/II/III (Taurus), to Class III (BPMG), to main sequence (the field) populations can explain the derived quantities.

\begin{figure}
\begin{center}
\includegraphics[width=0.45\textwidth]{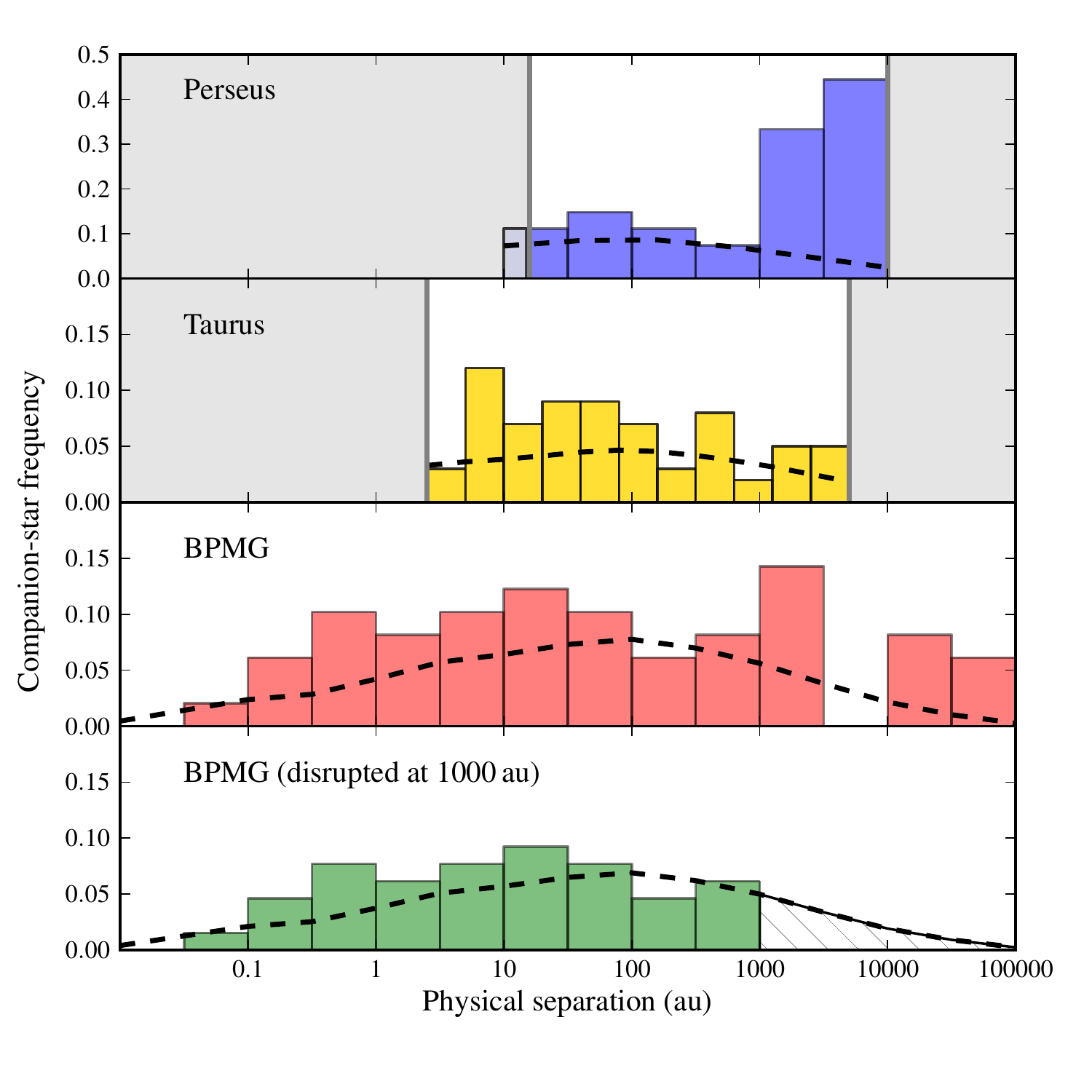}
\caption{Physical separation versus CSF for four different samples.  {\it Top:} Perseus (15-10,000\,au, \citealt{Tobin2016}), {\it second:} Taurus (3-5000\,au, \citealt{Kraus2011}), {\it third:} The BPMG sample, {\it bottom:} The BPMG sample, disrupted at 1000\,au.  The log-normal field distribution derived in \citet{Raghavan2010} is over-plotted (black dotted line) in all panels.  Grey shaded areas represent parameter spaces not probed.}
\label{fig:binary_seps}
\end{center}
\end{figure}

The top panel of Figure~\ref{fig:binary_seps} shows the {\it CSF} as a function of physical separation for Class 0 objects in Perseus \citep{Tobin2016}. {We chose the \cite{Tobin2016} sample as our main comparison to Class 0 sources as opposed to \cite{Chen2013} because of its completeness and focus on one specific region, the Perseus region. } The main limitation of the observations of Perseus is the angular resolution (0.\arcsec 065, $\approx$15\,au, grey line).  However, we would not expect to find components smaller than this separation due to the opacity limit, where the compression becomes approximately adiabatic, of the initial cores ($\approx$5\,au, \citealt{Larson1969}).  If components have undergone significant dynamical evolution at this stage already (as suggested by \citealt{Reipurth2010}) and are brought within this limit they would most likely have merged.  Additionally systems with separations greater than $\sim$5,000-10,000\,au at these young ages, are unlikely primordial (originating from different birth sites).  We cannot assess whether components are gravitationally bound however, as in \cite{Tobin2016}, we treat all identified components within 10,000\,au as bound systems.  The second panel shows the distribution for the Taurus region (3-5000\,au).  We see that there is a wealth of systems with separations $<$10\,au and the overall number of multiple systems is still very high.

{
\begin{table}
\tiny
\caption{A summary of the type and fraction of multiple systems samples analysed in this work. S:B:T:Qu:Qi:Se represent system components from Equation~\ref{eq:mf}.}
\begin{tabular}{p{1.95cm} p{0.3cm} p{1.5cm} p{0.3cm} p{1.1cm} p{1.1cm}}
\hline\hline\\
Sample & Ref. & S:B:T:Qu:Qi:Se & Total & {\it MF}$^a$ & {\it CSF}$^a$ \\[1ex]
\hline\\
Class 0 regions & 1 & 36:43:15:6:0:0 & 33 & 0.64$^{+0.09}_{-0.05}$ & 0.91$^{+0.15}_{-0.09}$ \\[1.3ex]
Perseus & 2 & 43:23:17:7:7:3 & 30 & 0.57$^{+0.06}_{-0.11}$ & 1.20$^{+0.27}_{-0.32}$ \\[1ex]
Taurus & 3 & 41:43:10:3:0:1 & 117 & 0.59$^{+0.04}_{-0.04}$ & 0.78$^{+0.09}_{-0.07}$ \\[1ex]
BPMG  & 4 & 35:35:24:6:0:0 & 49 & 0.65$^{+0.06}_{-0.02}$ & 1.02$^{+0.14}_{-0.06}$\\[1ex]
BPMG (10,000\,au) & 4 & 44:42:11:3:0:0 & 57 & 0.56$^{+0.07}_{-0.05}$ & 0.73$^{+0.12}_{-0.08}$\\[1ex]
BPMG (1000\,au) & 4 & 52:38:10:0:0:0 & 63 & 0.48$^{+0.06}_{-0.07}$ & 0.58$^{+0.08}_{-0.09}$\\[1ex]
The field & 5  & 56:33:8:3:0:0 & 454 & 0.44$^{+0.01}_{-0.02}$ & 0.58$^{+0.01}_{-0.04}$\\[1ex]
The field & 6 & 54:33:8:4:1:0 & 4847 & 0.46$^{+0.01}_{-0.01}$ & 0.65$^{+0.01}_{-0.01}$\\[1ex]
\hline\\[-0.1ex]
\end{tabular}
{\footnotesize $^a${$1 \sigma$ confidence intervals from 1000 bootstrapped samples. Note, bootstrapped uncertainties are indistinguishable from those calculated using binomial statistics for {\it MF} values in the majority of cases.
}}\\
{\footnotesize 1: \cite{Chen2013}, 2: \cite{Tobin2016}, 3: \cite{Kraus2011}, 4: This work, 5: \cite{Raghavan2010}, 6: \cite{Tokovinin2014b}}
\label{tab:mf_summary}
\end{table}
}

For comparison with the BPMG (middle panel of Figure~\ref{fig:binary_seps}) we assume all wide systems are primordial.
There is another mechanism to produce wide (non-primordial) systems \citep{Kouwenhoven2010}, which is discussed in Section~\ref{sec:remarks}.  At the same time as there being a wealth of systems with separations $<$10\,au in our sample there are also many objects with large separations (1000-100,000\,au).  This {\it stretching} of the initial separation distribution to both smaller separations and larger separations implies intense dynamical evolution.

\subsection{The role of unfolding higher-order systems}

\cite{Reipurth2012} described a potential physical mechanism for the formation of wide and very wide binaries ($\geq$1,000\,au); through the dynamical {\it unfolding} of primordial triple systems. 
The mechanism predicts:{
\begin{itemize}
\item A preference for wide binaries to be in triple systems
\item These systems will have high ($>$0.9) eccentricities.
\item The majority of wide components will be low-mass (these systems are likely unstable and therefore their lifetime is $<$100\,Myr)
\end{itemize}}

We note that although there should be a preference for wide binary systems to be in triple systems we would not expect to find all systems in this configuration due to previous mergers. 
Additionally that we cannot assess the eccentricity of the wide components.  This problem is discussed in more detail in Section~\ref{sec:remarks}.  

\subsubsection{Higher-order multiplicity in wide systems?}

We found that for systems with companions $>$1000\,au and $>$10,000\,au, 11 / 14 (0.79$^{+0.07}_{-0.14}$) and 6 / 7 (0.86$^{+0.05}_{-0.21}$) were part of triple or higher-order systems, respectively. 
In contrast, for systems with no components $>$1000\,au we found 4 / 19 (0.21$^{+0.12}_{-0.06}$) systems were higher-order systems.  {\it The fraction of higher-order multiple systems is clearly a function of the physical separation.}

Additionally we looked at the separations of the inner components of the 11 systems (8 triples, 3 quadruples, i.e. 14 inner components) with companions $>$1000\,au.  Of these 14, 11 had separations smaller than the inner peak identified of the separation distribution found in the Perseus region by \cite{Tobin2016}.  Again, this supports the theory of migration via the interaction of multiple components.

Although there is no evidence for higher-order multiplicity amongst the 3 binary systems (BPC 36,42,65) that have components $>$1000\,au, that does not necessarily negate the possibility, as few of the components have been observed with AO-imaging and have multi-epoch radial velocity data, see Table~\ref{tab:sample_targets}.

\subsubsection{Masses and mass-ratios}

We studied the mass distribution within the 10 higher-order multiple systems with wide components.  
For the triple systems we compared the mass-ratios of the inner binaries and outer components.  In the case of the quadruple systems we compared the mass-ratios of each sub-system and each component with that of the primary.  In neither case we found a significant relationship between the system configuration and absolute mass nor the mass-ratio. Our derived mass-ratio distribution was statistically similar to a power-law with $\gamma=0$ i.e. a flat distribution in the range 0.1-1.0.

\subsection{Disruption of higher-order systems}
\label{subsec:disrupt_stability}
 
In the framework of \cite{Reipurth2012} we estimated the survival rate of the identified triple systems, given their mass distribution.

\begin{figure}
\begin{center}
\includegraphics[width=0.45\textwidth]{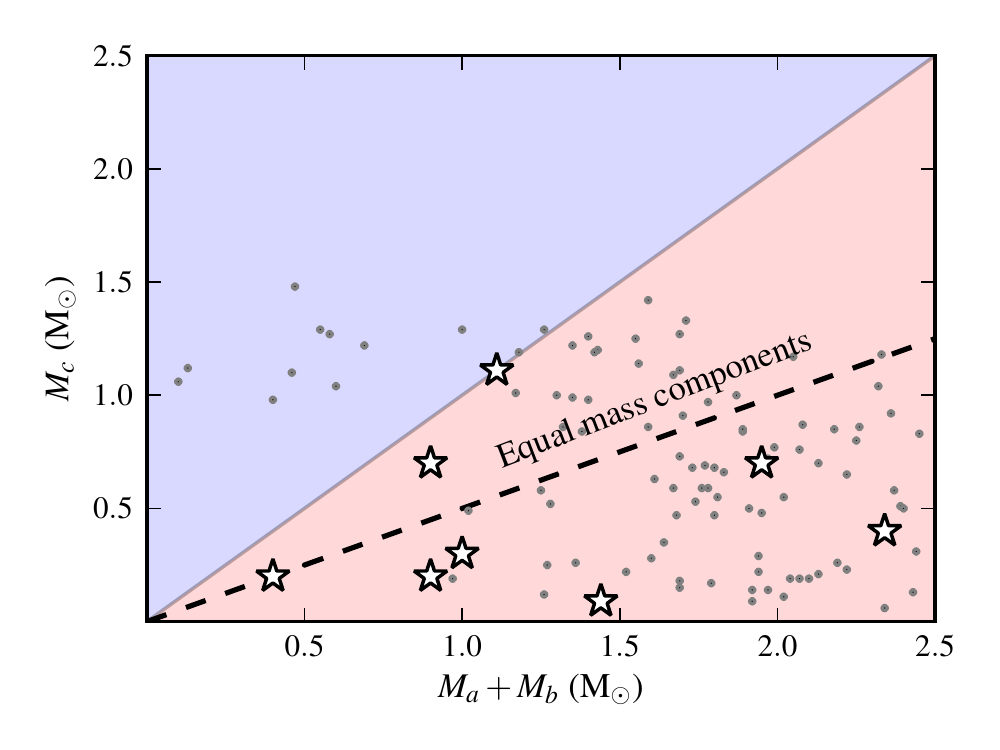}
\caption{Inner binary ($M_a+M_b$) versus outer component ($M_c$) masses for BPMG (stars) and field (\citet{Tokovinin2014b}, grey dots) wide systems. The blue and red shaded areas highlight the dominant single and binary regimes, respectively. The dotted line describes equal mass component systems (Ma = Mb = Mc).}
\label{fig:mass_sum}
\end{center}
\end{figure}

Similar to Figure~3 of \cite{Reipurth2012} we compared the mass sum of the inner binary ($M_a+M_b$) to the mass of the outer companion ($M_c$) for wide triple systems. We show our results in Figure~\ref{fig:mass_sum}: star markers are BPMG triple systems with at least one companion $>$1000\,au, grey dots are field stars \citep{Tokovinin2014b} meeting the same criterion.  Firstly we see that all BPMG systems are in the {\it dominant binary} (red shaded area) regime.  Secondly, we see that 5 / 8 of these systems occupy a parameter space mostly uninhabited by very wide field systems.   \cite{Reipurth2012} show that most systems in this parameter space  are unstable and are disrupted between the ages of 10-100\,Myr.
Therefore, our results imply that the majority of wide triple systems in the BPMG are in the process of disintegrating/decaying without external influence. 

To investigate if the disruption of higher-order multiple systems in our sample could ``bridge-the-gap" between young populations rich in multiplicity and the older processed field population we artificially disrupted our higher-order multiple systems at two physical separation limits (1000\,au and 10,000\,au)\footnote{The multiplicity properties derived using a disruption separation anywhere between 3000-10,000\,au are unaffected.}.
 The first limit is a somewhat arbitrary definition of a wide binary, the second is an approximation for the size of a hydrostatic clump.
 We then calculated the {\it MF} and {\it CSF}, considering systems with components wider than these limits as {\it separate} systems.
The results are shown in Table~\ref{tab:mf_summary}.  Considering the limit at 10,000\,au, the {\it MF} and {\it CSF} values are within 2\,$\sigma$ and 1\,$\sigma$, respectively, of the field population (higher in value).  For the limit at 1000\,au both the {\it MF} and {\it CSF} values are within 1$\sigma$ ({ the {\it CSF} }now lower in value).  In the same line, the {\it MF} and {\it CSF} values of Taurus are compatible with the disruption of systems at 10,000\,au in the BPMG sample.  
This suggests that the majority of primordial systems with separations somewhere between 1000-10,000\,au and beyond have been destroyed or decayed in older populations.

\begin{figure*}
\begin{center}
\includegraphics[width=0.9\textwidth]{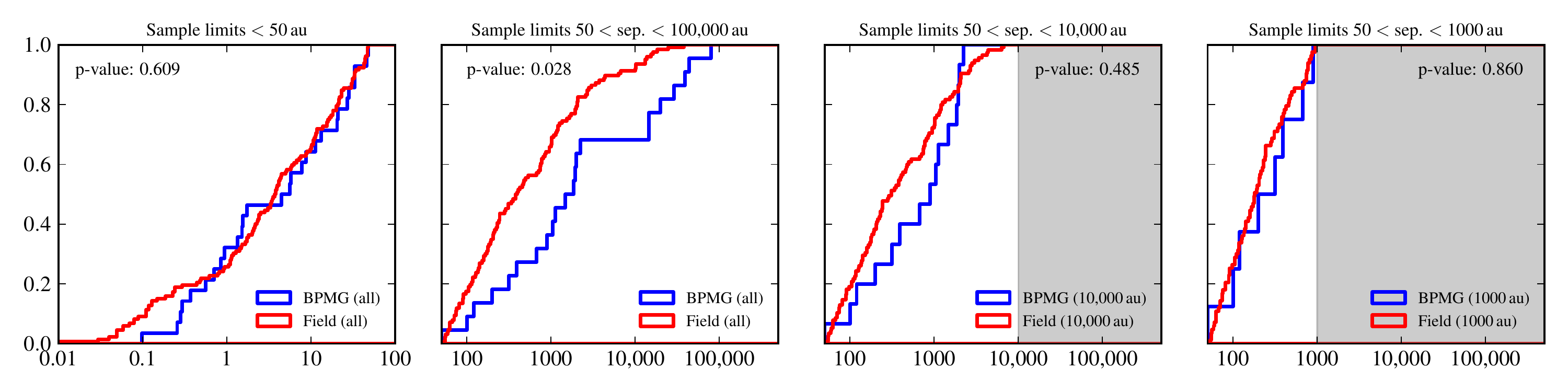}
\vspace{-0.2cm}
\caption{Cumulative separation distributions for four different samples.  Labels above each plot indicate the physical separation range.  The p-value calculated from the Kolmogorov-Smirnov test is shown in each panel.  Field systems from from \citet{Raghavan2010}.}
\label{fig:ks_comparison}
\end{center}
\end{figure*}

We also compared the {\it CSF}, as a function of physical separation distribution, for: the original BPMG sample, the BPMG sample disrupted at 1000\,au and 10,000\,au and the field.  
The results are shown in Figure~\ref{fig:ks_comparison}.  We split our samples at the mean of the field distribution ($\approx$50\,au) and compared the different populations.  
We would expect the distribution of inner separations to be very similar already at the age of the BPMG, as demonstrated in \cite{Elliott2014, Elliott2015}.{  In the left panel of Figure~\ref{fig:ks_comparison} we show that the distributions are indeed very similar.  The apparent difference at $\approx$0.3\,au is most likely an artefact of the assignment of physical separation values to spectroscopic systems currently without orbital solutions.}

We can reject the null hypothesis when comparing the field with the original BPMG sample (p-value 0.028, second panel of Figure~\ref{fig:ks_comparison}) for separations $>$50\,au, i.e. these distributions are not likely to be realisations of the same parent distribution.  However, if we compare the BPMG sample disrupted at 1000 and 10,000\,au ($3^{\rm rd}$ and $4^{\rm th}$ panel of Figure~\ref{fig:ks_comparison}) to the field, we calculate p-values of 0.860 and 0.485, respectively (non rejection).

The recent work of \cite{Parker2014b} suggested that the primordial binary population is similar to the population we observe today and has undergone little dynamical processing. 
Along with many previous works, \cite{Parker2014b} have focussed on binaries as opposed to higher-order multiple systems. However, our results indicate that the disruption of primordial higher-order systems significantly shapes older, more processed populations, such as the field.
  We outline a simple model for the contribution of unfolding triple systems to the separation distribution in a population in Section~\ref{sec:model}.

\section{A model for the dynamical evolution of triple systems}
\label{sec:model}

\begin{figure}
\begin{center}
\includegraphics[width=0.49\textwidth]{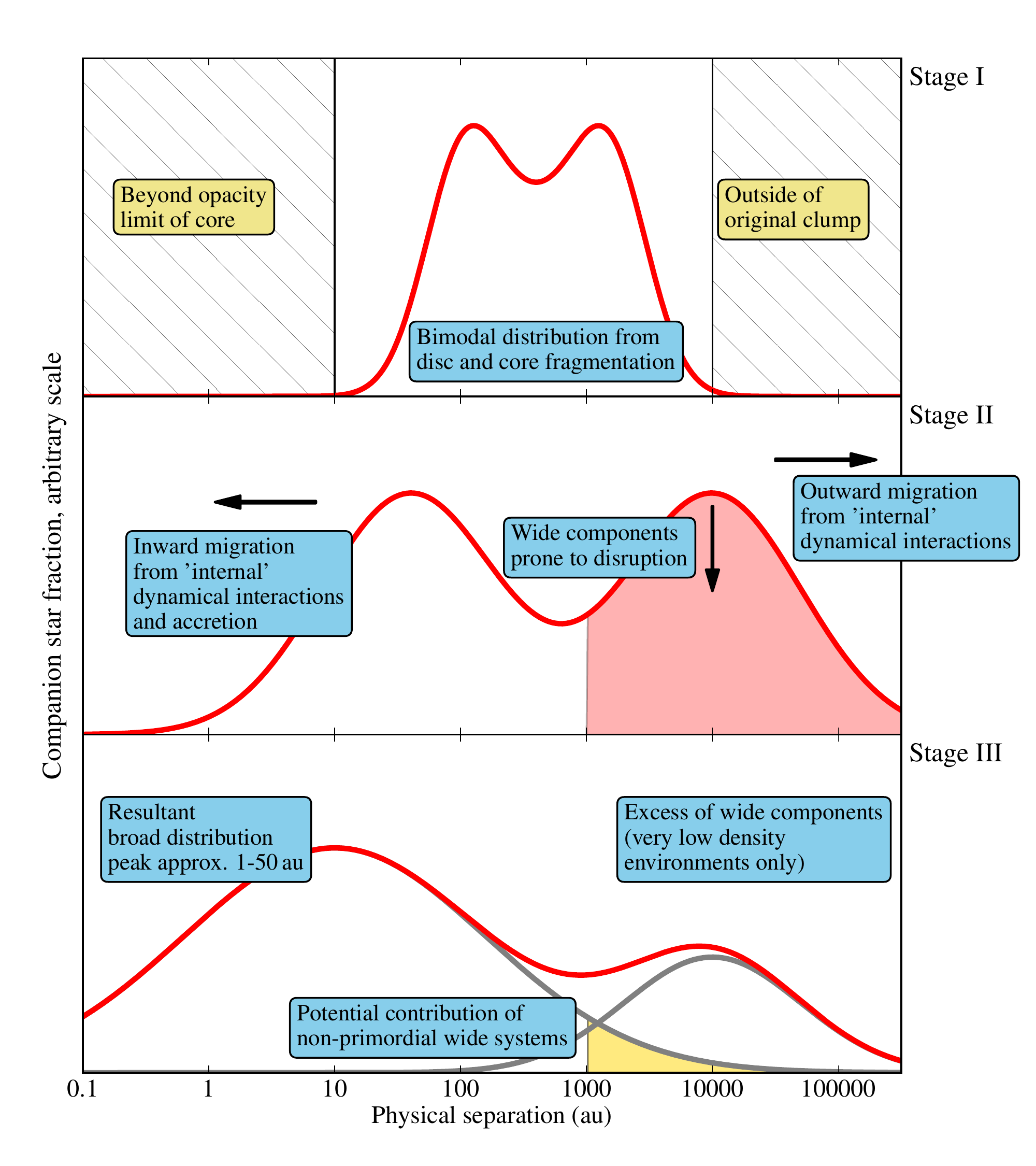}
\vspace{-0.2cm}
\caption{A simplified model of the dynamical evolution of separations within a population of primordial triple systems.  Regions of interest are annotated and processes are described in detail in the text (Section~\ref{sec:model}).}
\label{fig:triple_evolution}
\end{center}
\end{figure}

Below we collect together the physical mechanisms discussed in the previous sections, along with additional physics, to briefly outline the separation evolution of higher-order (in this case, specifically triple) systems in a population.  Our simplified model is shown in Figure~\ref{fig:triple_evolution} in 3 stages.\\

\noindent{\it Stage I}: We approximate the initial distribution of separations within clumps using the identified bimodal distribution of \cite{Tobin2016} (peaks; 75\ and 3000\,au). These peaks have been attributed with typical scales for disc ($\sim$100\,au, \citealt{Zhu2012}) and core ($\sim$1000\,au, \citealt{Offner2010}) fragmentation.  No multiple systems can inhabit the hatched region for separations $<$10\,au due to the size of the first Larson core (a pressure supported fragment) \citep[$\approx$5\,au, ][]{Larson1969}.  The second hatched region beyond $\approx$5000-10,000\,au represents the initial size of the hydrostatic clump \citep{Benson1989, Motte1998, Pineda2011}, i.e. by definition a primordial multiple system at this stage cannot have components beyond this separation.\\

\noindent{\it Stage II}: The inner peak both broadens and shifts to closer separations through the exchange of angular momentum with the outer component (components are now in a hierarchical configuration, although potentially unstable, as \citealt{Reipurth2012}). This is supported by the conservation of the {\it MF} and {\it CSF} quantities in larger separation ranges for Class 0 sources (15-10,000\,au, \citealt{Tobin2016}), Class I/II/III sources (3-5000\,au, \citealt{Kraus2011})
and Class III sources ($\approx$0.01-100,000\,au, this work).  Additional migration to closer separations can also occur through gas-induced orbital decay \citep{Korntreff2012, Bate2012}. Concurrently, through angular momentum exchanges, the outer components move to wider separations.  These wide components (red shaded area) are weakly bound, and some will be in the process of decaying (so-called {\it unfolding}: \citealt{Reipurth2012}) without any external influence.  Others may be dynamically disrupted / destroyed by external dynamics i.e. stellar encounters within a cluster.  Our analysis in Section~\ref{subsec:disrupt_stability} indicates that most systems are in the process of decaying without external influence. However, as this process is still taking place at the age of the BPMG, there is still a significant population of wide systems. \\

\noindent{\it Stage III}: In loose low-density environments, such as the BPMG, the resultant distribution is double peaked.  The inner peak is at a smaller separation than the original 75\,au ($\approx$4\,au for the higher-order multiple systems in our sample).  The outer peak ($\approx$8000\,au in our sample) is smaller as some components have unfolded and decayed completely, reducing the number of triples. However, a significant population still remains, due to the lack of external dynamical perturbations.  We have highlighted the region (gold shaded area) where non-primordial wide multiple systems can be formed during the cluster dissolution phase (\citealt{Kouwenhoven2010}, discussed in Section~\ref{sec:remarks}). In the case of higher-density and older ($\approx$100\,Myr) environments, the secondary peak is mostly destroyed and the contribution of primordial triple systems to the overall distribution is small/negligible.

\section{Discussion and limitations}
\label{sec:remarks}

We have shown that the dynamical evolution of primordial higher-order multiple systems can reconcile the multiplicity properties of Perseus, Taurus, BPMG and the field. Below we discuss the current limitations and prospects.

The mechanism of \cite{Kouwenhoven2010} is a potential route to produce wide, non-primordial binaries in the cluster dissolution phase.  However, this mechanism predicts only one or two wide binaries ($>$1000\,au) in a population of the size studied here, compared to the 14 identified.  Additionally we have shown how the decay of primordial higher-order systems can ``bridge-the-gap" between young and old populations.  We do not claim the mechanism of \cite{Kouwenhoven2010} is ineffective, rather that it is not the dominant mechanism in our sample.  { However, one also has to consider that binary formation and destruction is very stochastic.  N-body simulations of clusters with very similar initial conditions can produce resultant distributions with different forms \citep{Parker2012}.  Therefore, we must conduct similar analysis on other nearby young moving groups in order to see if those populations also have similar multiplicity properties to the BPMG.  }

A parameter that would help to determine the formation mechanism of wide multiple systems is their eccentricity ($e$).  { A natural consequence of the \cite{Reipurth2012} mechanism is a distribution peaked at high eccentricities. Whereas the mechanism of \cite{Kouwenhoven2010} predicts a flatter eccentricity distribution, see their Figure~10.}  
\cite{Tokovinin2016} recently showed that wide binaries in the field do not have a thermal eccentricity distribution and therefore are unlikely products of either formation scenario.  
  Given the predicted periods of very wide systems ($\gtrsim$\,Myr) determining their eccentricities is extremely challenging.  However, the {\sc gaia} mission \citep{Lindegren2008} will provide extremely accurate astrometry (7-25\,$\mu$as) that can help.   Face-on multiple systems of total mass 1\,M$_\odot$, at 40\,pc, and with 20,000\,au separation, will have detectable movement at 3$\sigma$ level using {\sc gaia} astrometry within 1\,yr. 
  
The work of \cite{Reipurth2012} used only three bodies within the original clump to study the formation of wide binaries.  However, the initial multiplicity is likely to be a spectrum, which indeed is observed in young regions (\citealt{Chen2013}, \citealt{Tobin2016}) where clumps contain up to six components.  Higher-order systems would be responsible for an even more chaotic start for components within the clump, likely resulting in more ejections.  
Further studies of nearby young regions with instruments such as ALMA and the VLA will provide further insight into the initial distributions of hydrostatic cores. These can then be used as realistic starting points of N-body simulations and {\it evolved} into older dynamically processed populations.

\section{Conclusions}
\label{sec:conclusions}

\begin{itemize}
\item Similar {\it CSF} values between the very young sources of Perseus and the BPMG can be explained by dynamical evolution of the separation distribution of multiple systems.
\item The high fraction of wide ($>$1000\,au, 11 / 14) and very wide ($>$10,000\,au, 6 / 7) systems in higher-order configurations supports the mechanism of \cite{Reipurth2012}.
\item There is no significant mass-ratio trend among inner and outer components in wide higher-order systems.
\item The separation distribution of the field and the BPMG (for separations $>$50\,au) are not realisations of the same parent distribution (p-value 0.03).
\item The decay of primordial higher-order (triple and higher-order) systems can link multiplicity in young regions to dynamically processed populations such as the field, in opposition to the more static framework of \cite{Parker2014b}.
\item The abundance and configuration of the wide systems in the BPMG favours the \cite{Reipurth2012} mechanism over that of \cite{Kouwenhoven2010}.
\end{itemize}

\section*{Acknowledgements}
PE would like to thank Bo Reipurth, Gaspard Duch\^ene and Richard Parker for discussions and comments that helped to improve this paper. AB acknowledges financial support from the Proyecto Fondecyt Iniciaci\'on 11140572


\bibliographystyle{mnras}
\bibliography{biblio1_orig}

\bsp	
\label{lastpage}
\end{document}